%% file: vdwl.tex
\documentclass[aps,prl,twocolumn,superscriptaddress,showpacs,floatfix]{revtex4}
\usepackage{color,amssymb,graphicx}
\newcommand{\beq}{\begin{equation}}
\newcommand{\eeq}{\end{equation}}
\newcommand{\bea}{\begin{eqnarray}}
\newcommand{\eea}{\end{eqnarray}}
\newcommand{\bean}{\begin{eqnarray*}}
\newcommand{\eean}{\end{eqnarray*}}
\newcommand{\bei}{\begin{itemize}}
\newcommand{\eei}{\end{itemize}}
\newcommand{\ben}{\begin{enumeration}}
\newcommand{\een}{\end{enumeration}}
\newcommand{\nn}{\nonumber}

\definecolor{darkorange}{rgb}{.6,.2,.0}

\definecolor{darkgreen}{rgb}{0.0,0.7,0.0}

\begin{document}

\title{Computational Consequences of Neglected First-Order 
van der Waals Forces}

\author{Kevin Cahill}
\email{cahill@unm.edu}
\affiliation{Department of Physics and Astronomy, 
University of New Mexico, Albuquerque, NM 87131}
\affiliation{Center for Molecular Modeling, 
Center for Information Technology, 
National Institutes of Health, 
Bethesda, Maryland 20892-5624}
\author{V.~Adrian Parsegian}
\email{aparsegi@helix.nih.gov}
\affiliation{National Institute of Child Health
and Human Development, National Institutes of Health, 
Bethesda, Maryland 20892-0924}

\date{\today}

\begin{abstract}
We have computed the widely neglected 
first-order interaction between neutral atoms.
At interatomic separations typical of condensed media, 
it is nearly equal to the \(-1/r^6\)
second-order London energy inferred from interactions in gases.
Our results, without the exchange forces
that lead to covalent bonding, suggest that the quality
of non-bonding attraction between neutral atoms of molecules 
in condensed media differs from the
\(-1/r^6\) form usually ascribed to it.  
If we add first-order and all second-order terms, we obtain energies
nearly five times the \(-1/r^6\) London energies which
dominate only at the atomic separations found in gases.
For computation, we propose a practical,
accurate form of energy to replace the qualitatively
inaccurate Lennard-Jones and harmonic forms
casually assumed 
to hold at the interatomic separations 
found in condensed media.
\end{abstract}


\maketitle

Does the van der Waals potential energy 
between two neutral atoms
fall off as \(\mbox{} - 1/r^6\) at 
separations of a few \AA?
Is the widely neglected first-order van der Waals potential 
zero at the interaction distances within condensed media?
Should one use a Lennard-Jones 6-12 potential 
in computer simulations?
Because it is no longer necessary to sacrifice accuracy 
for computational convenience, 
we numerically re-examine these old questions, 
find that common practice leads to qualitative error, 
and propose an accurate but practical alternative 
to the mathematically convenient but potentially  
misleading Lennard-Jones (L-J) and harmonic forms.
\par
Motivated by an intriguing footnote 
in Landau and Lifshitz~\cite{Landau1976},
we compute the first-order van der Waals potential 
for two hydrogen atoms, without the exchange interaction 
that creates covalent bonds.  It is nearly equal to 
the $1/r^6$ second-order London energy found from
the same wave functions when evaluated 
over a range of separations where both 
exceed $kT$ at room temperature. 
We then compute the potential energy
to second-order, obtaining an expression
that reassuringly reproduces 
the expected dominance of the \(1/r^6\)
London term as \(r \to \infty\)\@.
The net result at separations 
\(3 \, a_0 < r < 5 \, a_0\) 
(Bohr radius \(a_0\)) 
is an energy
4 to 5 times that expected from a London interaction, 
which emerges to dominate only in the 
\(r > 10 \, a_0\) limit. 
\par
We next examine the additivity of 
the interactions of several particles, 
usually taken to be additive 
in first and second order; 
the Axilrod-Teller-Muto force~\cite{Axilrod1943,Muto1943} 
is a third-order effect.  In fact, energies
add in first-order perturbation theory but not in second order
unless one neglects first-order terms. 
\par
Constructed to fit the measured equilibrium separation $r_s$ 
and the measured condensation energy energy $V(r_s)$ 
as well as to have the large-\(r\) asymptotic form $1/r^6$\@, 
the Lennard-Jones potential is too hard for \(r < r_s\)
and varies incorrectly for \(r > r_s\)\@.
Harmonic potentials work only near energy minima.
We propose a simple, phenomenological potential 
that can fit spectroscopic data and
that behaves appropriately at all distances.
\par
For our calculations, we take
the hamiltonian for two interacting hydrogen atoms to be
\(H = H_0 + W\)
in which \(H_0\) is a sum of two 
isolated-hydrogen-atom hamiltonians and
the perturbation \(W\) is a function
of the distance \( r \) between the two protons
\beq
W = \frac{e^2}{r} +
\frac{e^2}{|\vec r + \vec r_2 - \vec r_1|}
- \frac{e^2}{|\vec r + \vec r_2|}
- \frac{e^2}{|\vec r - \vec r_1|} .
\label{W}
\eeq
Here \(r = |\vec r|\) and 
\(e\) is the charge of the electron
in units with \(\alpha = e^2/(\hbar c) \approx 1/137\)\@.
The first-order correction to the
energy is 
\(\Delta E_1(r) = \langle 0 | W | 0 \rangle\)
in which the unperturbed ground state
\(| 0 \rangle = |100,100\rangle\) 
is two \(1s\) states separated by a distance \(r\)\@.
Its wave function is
\(
\langle \vec r_1, \vec r_2 | 0 \rangle =
(1/\pi a_0^3) \, \exp(-r_1/a_0 \, -r_2/a_0) 
\)
where \(a_0 = \hbar^2/(me^2)\) is the Bohr
radius, \( r_1 = |\vec r_1|\),
and \( r_2 = |\vec r_2|\)\@.
Integration of \( \langle 0 | W | 0 \rangle \)
yields
\beq
\Delta E_1(r) = mc^2 \alpha^2 \, e^{-2r/a_0}\,
\left( \frac{a_0}{r} + \frac{5}{8} 
- \frac{3r}{4a_0} - \frac{r^2}{6a_0^2} \right). 
\label{Delta E_1}
\eeq
London and others~\cite{Eisenschitz1930,London1930,Pauling1935} 
applied second-order perturbation theory to the
potential (\ref{W}), expanded for large \(r\), and got
\( \Delta E_L(r) = - 6.50 \, mc^2 \alpha^2\, a_0^6 / r^6 \)\@.
\par
\begin{figure}
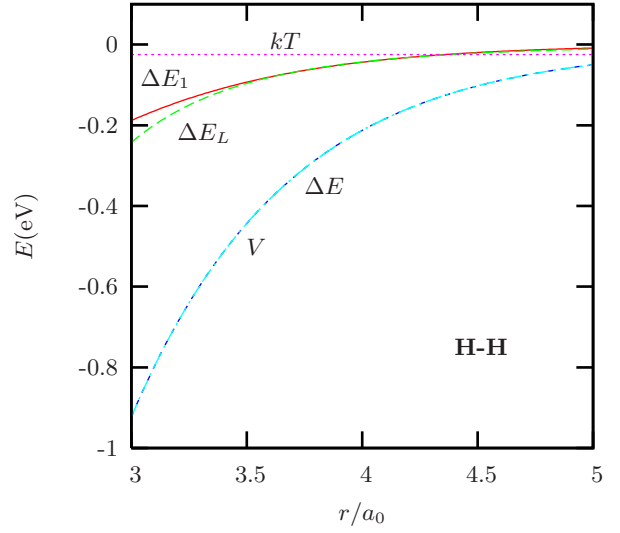

\centering
\input E2
\caption{The first-order van der Waals energy
\(\Delta E_1(r)\), Eq.(\ref{Delta E_1}), 
(solid, red) is nearly equal to
the London result \(\Delta E_L(r)\) (long dashes, green)
when both are reliable and greater than \(kT\) (dotted, plum).
The change in the sum 
\(\Delta E(r) = \Delta E_1(r) + \Delta E_2(r) \)
(dashes, blue) as given by 
second-order perturbation theory, Eq.(\ref{estimate}), 
is 4 to 5 times greater than the London term 
for \( 3 a_0 < r < 5 a_0 \)\@.
The potential of Eq.(\ref{phenpot})
with \(a=336\) eV, \(b=1.97 a_0^{-1}\),
\(c = 0.47 a_0^{-1}\), \(d = 397 a_0^6\) eV,
and \(e = 14093 a_0^{12}\) (dashes, light blue)
overlaps \(\Delta E(r)\)\@.}
\label{e2}
\end{figure}

Fig.~\ref{e2} shows that
the two energies \(\Delta E_1(r)\) and \(\Delta E_L(r)\)
are nearly equal 
for \(3 a_0 < r < 5 a_0\), 
where both are reliable and greater than \(kT = 1/40\) eV\@.
Because \(\Delta E_1(r)\) is the first-order result,
and \(\Delta E_L(r)\) is an approximation
to second-order perturbation theory,
the two can be added.
That is, to (approximate) second order in the perturbation \(W\)
and within the London approximation,
the van der Waals energy is 
\(
\Delta E_1(r) + \Delta E_L(r)
\),
different by about 100\% from the London-only result 
for $3 a_0 < r < 5 a_0$\@. 
Because \(\Delta E_1(r)\) decreases with \(r\)
as \(\exp(-2r/a_0)\), the London potential \(\Delta E_L(r)\)
dominates at large \(r\).  But it is not
until \(r = 5.78 \, a_0\) that \(|\Delta E_L(r)| >
2|\Delta E_1(r)|\)\@.
By then \(|\Delta E_L(r)| < kT/5\)\@.
\par
What about second order beyond $\Delta E_L(r)$\@?
The correction 
to the ground-state energy
due to the perturbation \(W\) is
\beq
\Delta E_2(r) = - \sum_{n \ne 0} \frac{|\langle 0 | W | n \rangle|^2}
{E_n - E_0}
\label{dE2}
\eeq
in which the sum is over all eigenstates
\(|n\rangle\) of the hamiltonian \(H_0\)
except \( | 0 \rangle \)\@.
If the first state to contribute to the sum
has energy \(E_1\), then
\( 0 < E_1 - E_0 \le E_n - E_0 \)
or 
\( - 1/(E_1 - E_0) \le - 1/(E_n - E_0) \)\@.
So the second-order change in the energy is bounded by
\beq
\mbox{} - \frac{1}{E_1 - E_0} 
\sum_{n \ne 0} |\langle 0 | W | n \rangle|^2
\le \Delta E_2(r) \le 0 .
\label{ne1}
\eeq
The first states to contribute to the sum
are those in which one of the electrons
is in the \(n=2\) level, 
so \(E_1 = -(1/2)\alpha^2 mc^2 ( 1 + 1/4)\)\@. 
But because the hydrogen wave functions
fall off as \( \exp(-r/(n \, a_0)) \),
at long distances the first states 
to contribute are those with both electrons
in the \(n=2\) level; they have 
\(E_1 = -(1/2)\alpha^2 mc^2 ( 1/4 + 1/4)\)\@. 
Without the energy denominators,
the sum over intermediate states
follows from completeness:
\beq
\sum_{n \ne 0} |\langle 0 | W | n \rangle|^2
= \langle 0 | W^2 | 0 \rangle
- \langle 0 | W | 0 \rangle^2 ,
\label{completeness}
\eeq
where \( \langle 0 | W | 0 \rangle \)
is the first-order result \( \Delta E_1(r) \),
Eq.(\ref{Delta E_1})\@.
\par
Evaluating the mean value of
\( W^2 \) involves the integral 
\( G(\vec x) = (2/\pi) \, \int \!\! d^3y \, 
(1/\vec y ^2) \, \exp(-2| \vec y - \vec x|)\), 
a function only of the
length \(x = |\vec x| = |\vec r|/a_0\)\@.
We find
\beq
G(x) = \frac{1}{x} \, 
\left[ (1 + 2x) \, e^{-2x} \, \mathrm{Ei}(2x) + 
( 1 - 2x) \, e^{2x} \, \mathrm{E_1}(2x) \right]
\label{g} 
\eeq
in which Ei and \( \mathrm{E_1} \) are exponential integrals.
\( G(x) \) contributes to
the integral \( J(x) \) 
\bea
J(x) & = & \frac{1}{4 x} \,
\int_0^\infty \!\! y \, G(y) \, \left[ 
\left( 1 + 2|x-y| \right) \, e^{-2|x-y|} \right. \nn \\
& & \qquad \qquad \left. \mbox{} - \left( 1 + 2(x+y) \right) 
\, e^{-2(x+y)} \right] \, dy
\label{J}
\eea
that represents the square
of the interaction of the
two electrons.
The product of electron-electron
and electron-proton energies is
described by the function \( K(x) \) 
\bea
K(x) & = & 
\left\{
4 \ln 2 - 3 + \frac{2\ln 2}{x} + 2x  \right. \nn \\
& & \left. \mbox{} - \left( 4 + \frac{2}{x} \right)
\left[ \, \mathrm{Ei}(2x) - \gamma - \ln(2x) \, \right] \right\} e^{-2x} \nn \\
& & \mbox{} + \left( 4 - \frac{2}{x} \right)
e^{2x} \left[ \, \mathrm{E_1}(2x) - \mathrm{E_1}(4x) \, \right]
\label{K}
\eea
in which \( \gamma = 0.5772\dots \) is the
Euler-Mascheroni constant.
In terms of \(\Delta E_1\), \(J\), and \(K\),
the second-order correction
\( \Delta E_2(r) \) to the energy at \(r = a_0 \, x \) 
becomes
\( - ( f \, a_0 /e^2 ) \, ( \langle 0 | W^2 | 0 \rangle
- \langle 0 | W | 0 \rangle^2 ) \) or 
\bea
\Delta E_2(r) & = & - \frac{f a_0}{e^2} \, \left\{
- \left( \Delta E_1(r) - \frac{e^2}{r} \right)^2 \right. \nn \\
& & \mbox{} + \frac{2e^4}{r^2} \, 
\left[ \, 1 - (1+ x )\,e^{-2x} \, \right]^2 
\label{estimate} \\
& & \mbox{} + \frac{e^4}{a_0^2} \, 
\left[ \, G(x) + J(x) + K(x) \, \right] \Biggr\}. \nn 
\eea
The fudge factor \( f = 1.083 = 6.499/6\) 
is the one used by Pauling and Beach~\cite{Beach1935} to 
estimate the effect of the energy denominators
in the sum Eq.(\ref{dE2})\@.
\par
To second order in the perturbation \(W\),
the change in the energy of two hydrogen atoms
at proton separation \( r \) is the sum
of the first-order term Eq.(\ref{Delta E_1}) and the
second-order term Eq.(\ref{estimate}),
\(
\Delta E(r) = \Delta E_1(r) + \Delta E_2(r) 
\label{DE}
\)\@.
Fig.~\ref{e2} shows
\( \Delta E(r) \) for
\( 3 \le r/a_0 \le 5\)
where \( \Delta E\) is never
even close to \( \Delta E_L\)\@.
(To model non-reactive atoms, 
we have neglected the exchange terms 
that cause covalent bonding and only 
increase \( | \Delta E(r) | \)\@.)
\par
What about additivity? 
Consider three hydrogen atoms
at \( \vec R_1 \),
\( \vec R_2 \), and \( \vec R_3 \)\@.
Again neglect exchange effects and
take the unperturbed ground state to be 
\beq
| 1, 2, 3 \rangle =
| 01, 02, 03 \rangle =
|1s,\vec R_1;1s,\vec R_2;1s,\vec R_3\rangle .
\label{RRR}
\eeq
Now the perturbation \( W \) is
\( W = W_{12} + W_{23} + W_{13} \)
in which the pair potential \( W_{ij} \) is
\bea
W_{ij} & = & \frac{e^2}{|\vec R_j - \vec R_i|} +
\frac{e^2}{|\vec R_j + \vec r_j - \vec R_i - \vec r_i|} \nn \\
& & \mbox{}
- \frac{e^2}{|\vec R_j + \vec r_j - \vec R_i |}
- \frac{e^2}{|\vec R_j - \vec R_i - \vec r_i |} .
\label{Wij}
\eea
To first order in the perturbation \( W, \) 
the change in the energy 
\(\Delta E_1 = \langle  1, 2, 3 |
\, W_{12} + W_{23} + W_{31} \,
| 1, 2, 3 \rangle \)
is additive:
\( \Delta E_1 = \langle 1, 2| \, W_{12} \, | 1, 2 \rangle 
+ \langle 2, 3 | \, W_{23} \, | 2, 3 \rangle  
+ \langle 1, 3 | \, W_{13} \, | 1, 3 \rangle \)\@.
\par
But at smaller separations the cross-terms do not vanish.
To second order in \( W \),
\(\Delta E_2\) is not quite additive.
The mean value of \( W_{12} \, W_{23} \), {\it e.g.,}
involves a non-zero sum over the intermediate states
\( | 1, n2, 3 \rangle = |01, n2, 03 \rangle \),
\bea
\lefteqn{\langle  1, 2, 3 |
 W_{12} \, W_{23} 
| 1, 2, 3 \rangle = } \nn \\
& & \sum_n \langle  1, 2, 3 |  W_{12} | 1, n2, 3 \rangle 
\langle  1, n2, 3 |  W_{23} | 1, 2, 3 \rangle \nn\\
& = & 
\sum_n \langle  1, 2 |  W_{12} | 1, n2 \rangle 
\langle n2, 3 |  W_{23} | 2, 3 \rangle \ne 0 .
\label{2NA}
\eea
Failure of additivity already occurs
at second order.
\par
In the usual treatment,
one expands
the perturbation \( W \) in
powers of \( 1/r \) keeping only terms as big as
\( 1/r^6 \) at large \( r \) and
neglects terms like Eq.(\ref{2NA}).
If \( \vec R_2 - \vec R_1 \) points in the
3-direction, then
the leading term in the matrix element 
\( \langle  1, 2 |  W_{12} | 1, n2 \rangle \)
involves \( (x_1 x_2 + y_1 y_2 - 2 z_1 z_2 )/| \vec R_2 - \vec R_1 |^3 \)
which vanishes because of the spherical symmetry
of the ground state of atom 1\@.
This neglect of the first-order van der Waals force
defers non-additivity 
to the third order of perturbation theory in the
Axilrod-Teller-Muto formulation~\cite{Axilrod1943,Muto1943}\@.
This deferral, however,
is probably inconsequential compared with 
the prior omission of purely additive first-order interactions.
\par
\begin{figure}
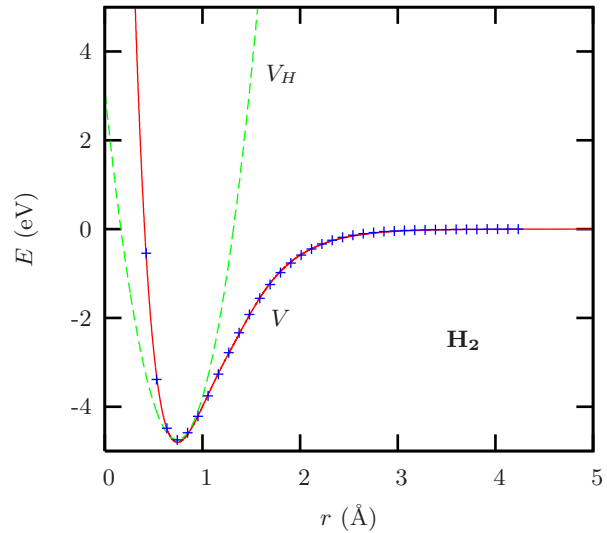

\centering
\input RKR
\caption{The phenomenological potential (\ref{phenpot}) 
with \(a = 53.8\) eV,
\(b = 2.99\) \AA\(^{-1}\),
\(c = 2.453\) \AA\(^{-1}\),
\(d = C_6 = 3.884\) eV\,\AA\(^6\),
and \(e = 47.6\) \AA\(^{12}\)
(solid, red) is finite,
fits the RKR spectral points for 
molecular hydrogen (crosses, blue),
and gives the correct London tail for \(r > 3\) \AA\@.
The harmonic potential \(V_H\) (dashes, green)
is accurate only near its minimum.}
\label{rkr}
\end{figure}
\begin{figure}
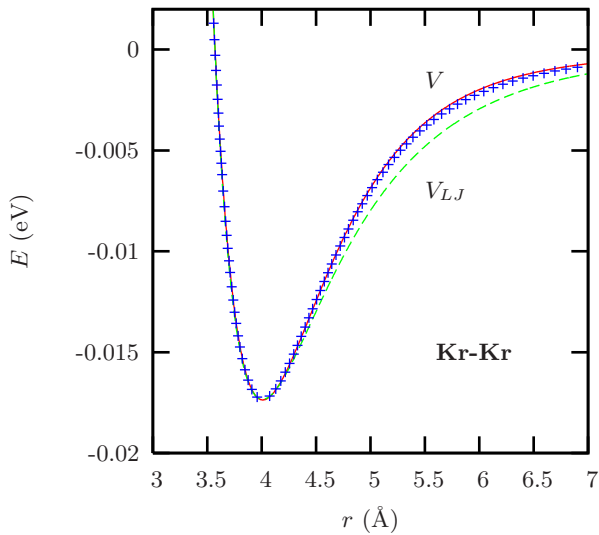

\centering
\input KR237
\caption{The potential \(V\) 
of Kr-Kr attraction (Eq.(\ref{phenpot}) 
with \(a = 2296\) eV,
\(b = 2.5136\) \AA\(^{-1}\),
\(c = 0.2467\) \AA\(^{-1}\),
\(d = 78.2146\) eV\,\AA\(^6\), and
\(e = 407923\) \AA\(^{12}\))
(solid, red) is finite, fits
the Kr\(_2\) Aziz points (crosses, blue),
and gives the correct London tail.
When matched at the minimum,
the Lennard-Jones form \(V_{LJ}\) 
(Eq.(\ref{LJ}) with \(r_s = 4.008\) \AA\ 
and \(V(r_s) = -0.017338\) eV) (dashes, green) 
is too low for \(r > 4\) \AA.}
\label{kr237}
\end{figure}
\begin{figure}
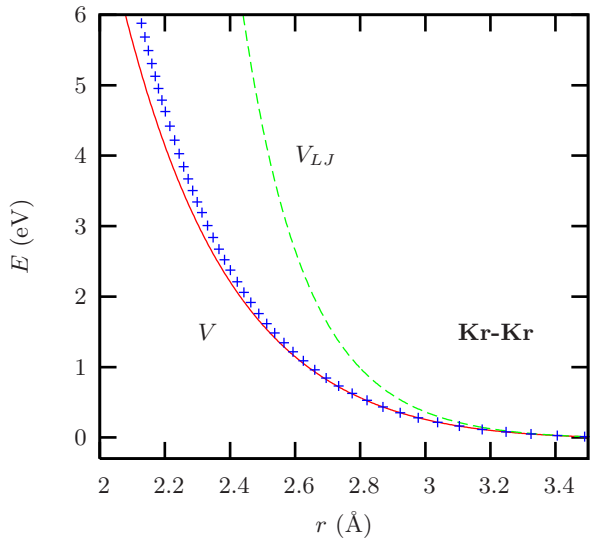

\centering
\input KR2
\caption{Positive potential \(V\) (Eq.(\ref{phenpot}) 
as in Fig.~(\ref{kr237})),
Kr\(_2\) Aziz points (crosses, blue),
L-J form \(V_{LJ}\) (Eq.(\ref{LJ})
as in Fig.~(\ref{kr237}))\@.}
\label{kr2}
\end{figure}
\par
Clearly there is room for improvement on the 
Lennard-Jones~\cite{Margenau1939,Pastor1984,Atlas1988,Halgren1992,
Meath1978,Buckingham1988,Ben-Amotz1990,Ben-Amotz1993,
Chalasinski1994,Jeziorski1994,vanLenthe1994,MacKerell1998,Choy2000,
MacKerell2001,Ben-Amotz2002,Persson2002}
and harmonic forms 
\bea
V_{LJ}(r) & = & |V(r_s)| \, \left[ \left( \frac{r_s}{r} \right)^{12}
- 2 \, \left( \frac{r_s}{r} \right)^6 \right]
\label{LJ} \\
V_H(r) & = & V(r_s) 
+ \frac{(r - r_s)^2}{2} \, 
\frac{d^2 V(r_s)}{dr^2} .
\label{Ha}
\eea
What to do?
We suggest using the form 
\beq
V(r) = a e^{-b \, r}\,( 1 - c \, r ) 
- \frac{d}{r^6 + e \, r^{-6}}.
\label{phenpot}
\eeq
The terms involving  \(a\), \(b\), and \(c\) 
appear in Eq.(\ref{Delta E_1}) for \(\Delta E_1(r)\),
were proposed by Rydberg~\cite{Rydberg1931}
to incorporate spectroscopic data,
but were largely ignored until 
recently~\cite{Ferrante1991,Murrell1984}.
The constant \(d = C_6\)
is the coefficient of the London tail.
The new term \( e \, r^{-6} \) cures
the London singularity;
as \(r \to 0\),
\(d/(r^6 + e \, r^{-6})\to
d \, r^6/e\) and
\( V(r) \to a\), finite.
But as \(r \to \infty\), \(V(r)\) 
approaches the London term,
\(V(r) \to - d/r^6 = -C_6/r^6\)\@.
\par
To test the ability of the potential \(V(r)\)
to represent covalent bonds,
we used Gnuplot~\cite{Gnuplot}
to choose \(a\), \(b\), \(c\), \(d\), and \(e\)
so as to fit Eq.(\ref{phenpot}) 
to the experimental potentials of
molecular H\(_2\), N\(_2\), and O\(_2\)
obtained from spectroscopic 
data~\cite{Weissman1963,Krupenie1977,Krupenie1972}
by the 
Rydberg~\cite{Rydberg1931}-Klein~\cite{Klein1932}-
Rees\cite{Rees1947} (RKR) method.
For H\(_2\), we set \(d\) equal to the London value,
\(d = C_6 = 6.5 \, mc^2 \, \alpha^2 \, a_0^6\)\@.
Fig.(\ref{rkr}) shows
that the potential of Eq.(\ref{phenpot}) (solid, red)
goes through the RKR points for H\(_2\) (crosses, blue),
while the harmonic potential (\ref{Ha}) 
(dashed, green)
is accurate only near its minimum and is 
dreadful for \(r \gg r_s\)\@. 
The fits of \(V(r)\) to the
N\(_2\) and O\(_2\) 
RKR points were equally good.
The potential \(V(r)\) of Eq.(\ref{phenpot}) 
represents covalent bonds better than
the harmonic potential \(V_H\)
of Eq.~(\ref{Ha}) (or \(V_{LJ}\))\@.
\par
Can \(V(r)\) represent 
weak noncovalent bonds?
Using Gnuplot, we fitted 
\(a\), \(b\), \(c\), and \(e\) in Eq.(\ref{phenpot})
to calculated He\(_2\)~\cite{Anderson2001}
or empirical Ar\(_2\) and Kr\(_2\)~\cite{Aziz1986} 
potentials, setting \(d\) equal to the 
respective London coefficient. 
Fig.(\ref{kr237}) shows that 
\(V(r)\) of Eq.(\ref{phenpot}) (solid, red) 
fits the empirical points~\cite{Aziz1986}  
for Kr\(_2\) (crosses, blue), 
but that the Lennard-Jones potential \(V_{LJ}\) 
of Eq.(\ref{LJ}) (dashes, green)
is too low for \( r > 4\)\AA\@.
Fig.~(\ref{kr2}) shows that \(V(r)\)
fits the empirical Kr\(_2\) points, 
but that \(V_{LJ}\) 
is much too hard for \( r < 4\)\AA\@.
The fits for He\(_2\) and Ar\(_2\)
were equally good.
Figs.(\ref{rkr}-\ref{kr2}) would look even better
with the London \(d\) treated as a free parameter.
The potential \(V(r)\) of Eq.(\ref{phenpot}) 
represents weak noncovalent bonds
better than \(V_{LJ}\)
or \(V_H\)\@.
\(V_{LJ}\) is unnaturally hard for \(r < r_s\);
it can wrongly reject
Monte Carlo moves that put atoms 
slightly too close,
frustrating M-C searches 
for the native states of macromolecules.
\par
The van der Waals potential 
is often as large in first order as 
in second.
The potential 
(\ref{phenpot})
may be useful in macromolecular simulations
that involve not only bonded 
but also non-bonded atoms.
Given the differences between
Eq.(\ref{phenpot}) and the popular
Lennard-Jones form, it would be worthwhile
to examine the consequences of
these differences in numerical
simulations.
\begin{acknowledgments}
Thanks to E.~Arriola and A.~Calle for pointing out
a typo in Eq.(\ref{J});
to S.\ Valone for RKR data;
to S.\ Atlas, C.\ Beckel,
B.\ Brooks, J.\ Cohen, K.\ Dill, D.\ Harries, G.\ Herling,
M.\ Hodoscek, R.\ Pastor, 
R.\ Podgornik, W.\ Saslow, and C.\ Schwieters for advice.
P.~J.\ Steinbach kindly hosted KC\ at NIH,
where we used the Biowulf computers.
\end{acknowledgments}

\bibliography{chem,cs,physics,vdw}

\end{document}

%% file: E2.tex
  \setlength{\unitlength}{0.1bp}%
\begin{picture}(2340,2073)(0,0)%
\put(1668,680){\makebox(0,0)[l]{\textbf{H-H}}}%
\put(972,1837){\makebox(0,0)[l]{\(kT\)}}%
\put(885,1061){\makebox(0,0)[l]{\(V\)}}%
\put(1103,1289){\makebox(0,0)[l]{\(\Delta E\)}}%
\put(624,1487){\makebox(0,0)[l]{\(\Delta E_L\)}}%
\put(485,1692){\makebox(0,0)[l]{\(\Delta E_1\)}}%
\put(1320,50){\makebox(0,0){\(r/a_0\)}}%
\put(100,1137){%
\special{ps: gsave currentpoint currentpoint translate
270 rotate neg exch neg exch translate}%
\makebox(0,0)[b]{\shortstack{\(E\)(eV)}}%
\special{ps: currentpoint grestore moveto}%
}%
\put(2190,200){\makebox(0,0){ 5}}%
\put(1755,200){\makebox(0,0){ 4.5}}%
\put(1320,200){\makebox(0,0){ 4}}%
\put(885,200){\makebox(0,0){ 3.5}}%
\put(450,200){\makebox(0,0){ 3}}%
\put(400,1822){\makebox(0,0)[r]{ 0}}%
\put(400,1517){\makebox(0,0)[r]{-0.2}}%
\put(400,1213){\makebox(0,0)[r]{-0.4}}%
\put(400,909){\makebox(0,0)[r]{-0.6}}%
\put(400,604){\makebox(0,0)[r]{-0.8}}%
\put(400,300){\makebox(0,0)[r]{-1}}%
\includegraphics{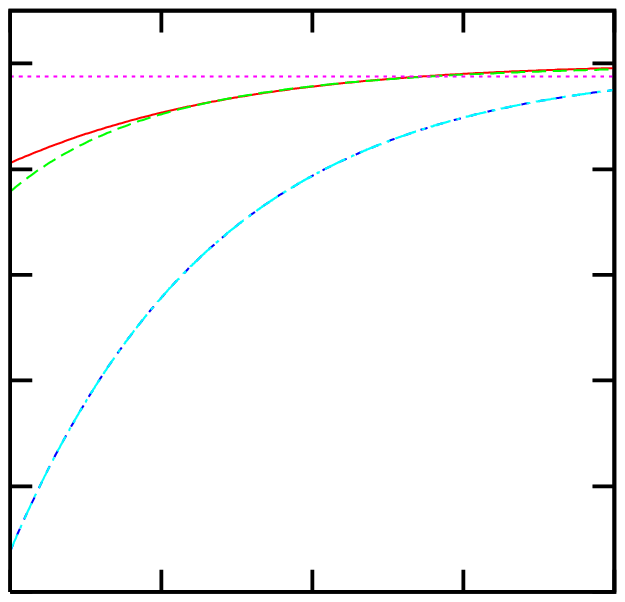}
\end{picture}

%% file: RKR.tex
  \setlength{\unitlength}{0.1bp}%
\begin{picture}(2340,2073)(0,0)%
\put(957,1723){\makebox(0,0)[l]{\(V_H\)}}%
\put(976,802){\makebox(0,0)[l]{\(V\)}}%
\put(1638,719){\makebox(0,0)[l]{\textbf{H}\(\mathbf{_2}\)}}%
\put(1270,50){\makebox(0,0){\(r\) (\AA)}}%
\put(100,1137){%
\special{ps: gsave currentpoint currentpoint translate
270 rotate neg exch neg exch translate}%
\makebox(0,0)[b]{\shortstack{\(E\) (eV)}}%
\special{ps: currentpoint grestore moveto}%
}%
\put(2190,200){\makebox(0,0){ 5}}%
\put(1822,200){\makebox(0,0){ 4}}%
\put(1454,200){\makebox(0,0){ 3}}%
\put(1086,200){\makebox(0,0){ 2}}%
\put(718,200){\makebox(0,0){ 1}}%
\put(350,200){\makebox(0,0){ 0}}%
\put(300,1807){\makebox(0,0)[r]{ 4}}%
\put(300,1472){\makebox(0,0)[r]{ 2}}%
\put(300,1137){\makebox(0,0)[r]{ 0}}%
\put(300,802){\makebox(0,0)[r]{-2}}%
\put(300,467){\makebox(0,0)[r]{-4}}%
\includegraphics{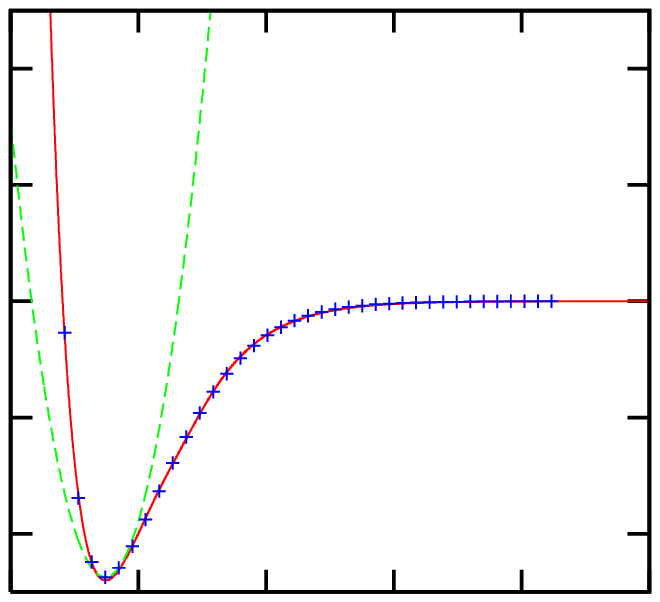}
\end{picture}

%% file: KR237.tex
  \setlength{\unitlength}{0.1bp}%
\begin{picture}(2340,2073)(0,0)%
\put(1575,1708){\makebox(0,0)[l]{\(V\)}}%
\put(1575,1289){\makebox(0,0)[l]{\(V_{LJ}\)}}%
\put(1616,680){\makebox(0,0)[l]{\textbf{Kr-Kr}}}%
\put(1370,50){\makebox(0,0){\(r\) (\AA)}}%
\put(100,1137){%
\special{ps: gsave currentpoint currentpoint translate
270 rotate neg exch neg exch translate}%
\makebox(0,0)[b]{\shortstack{\(E\) (eV)}}%
\special{ps: currentpoint grestore moveto}%
}%
\put(2190,200){\makebox(0,0){ 7}}%
\put(1985,200){\makebox(0,0){ 6.5}}%
\put(1780,200){\makebox(0,0){ 6}}%
\put(1575,200){\makebox(0,0){ 5.5}}%
\put(1370,200){\makebox(0,0){ 5}}%
\put(1165,200){\makebox(0,0){ 4.5}}%
\put(960,200){\makebox(0,0){ 4}}%
\put(755,200){\makebox(0,0){ 3.5}}%
\put(550,200){\makebox(0,0){ 3}}%
\put(500,1822){\makebox(0,0)[r]{ 0}}%
\put(500,1441){\makebox(0,0)[r]{-0.005}}%
\put(500,1061){\makebox(0,0)[r]{-0.01}}%
\put(500,680){\makebox(0,0)[r]{-0.015}}%
\put(500,300){\makebox(0,0)[r]{-0.02}}%
\includegraphics{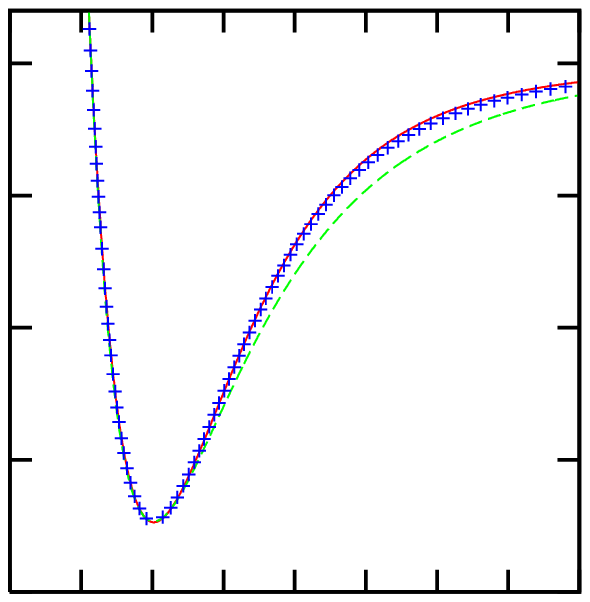}
\end{picture}

%% file: KR2.tex
  \setlength{\unitlength}{0.1bp}%
\begin{picture}(2340,2073)(0,0)%
\put(718,778){\makebox(0,0)[l]{\(V\)}}%
\put(1086,1443){\makebox(0,0)[l]{\(V_{LJ}\)}}%
\put(1699,778){\makebox(0,0)[l]{\textbf{Kr-Kr}}}%
\put(1270,50){\makebox(0,0){\(r\) (\AA)}}%
\put(100,1137){%
\special{ps: gsave currentpoint currentpoint translate
270 rotate neg exch neg exch translate}%
\makebox(0,0)[b]{\shortstack{\(E\) (eV)}}%
\special{ps: currentpoint grestore moveto}%
}%
\put(2067,200){\makebox(0,0){ 3.4}}%
\put(1822,200){\makebox(0,0){ 3.2}}%
\put(1577,200){\makebox(0,0){ 3}}%
\put(1331,200){\makebox(0,0){ 2.8}}%
\put(1086,200){\makebox(0,0){ 2.6}}%
\put(841,200){\makebox(0,0){ 2.4}}%
\put(595,200){\makebox(0,0){ 2.2}}%
\put(350,200){\makebox(0,0){ 2}}%
\put(300,1974){\makebox(0,0)[r]{ 6}}%
\put(300,1708){\makebox(0,0)[r]{ 5}}%
\put(300,1443){\makebox(0,0)[r]{ 4}}%
\put(300,1177){\makebox(0,0)[r]{ 3}}%
\put(300,911){\makebox(0,0)[r]{ 2}}%
\put(300,645){\makebox(0,0)[r]{ 1}}%
\put(300,380){\makebox(0,0)[r]{ 0}}%
\includegraphics{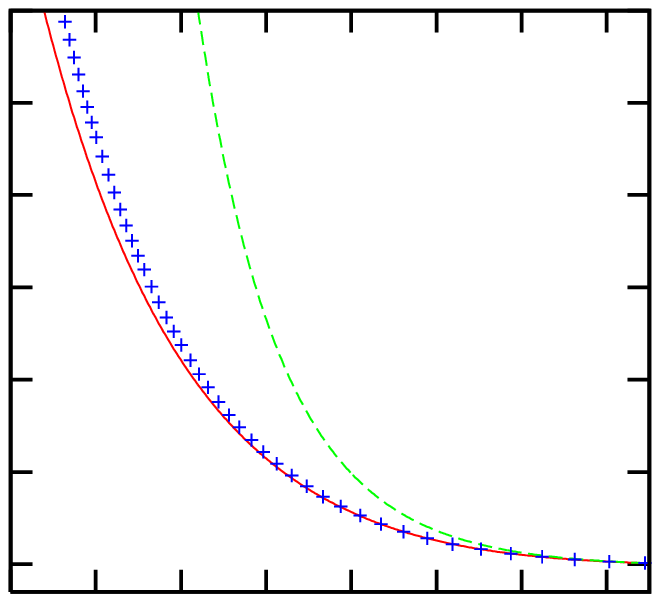}
\end{picture}